\documentclass[conference]{IEEEtran}

\usepackage[utf8]{inputenc}
\usepackage{cite}
\usepackage{amsmath,amssymb,amsfonts}
\usepackage{algorithmic}
\usepackage{graphicx}
\usepackage{textcomp}
\usepackage{xcolor}
\usepackage{booktabs}
\usepackage{multirow}
\usepackage{placeins}

\usepackage{threeparttable}  
\usepackage[normalem]{ulem}

\def\BibTeX{{\rm B\kern-.05em{\sc i\kern-.025em b}\kern-.08em
    T\kern-.1667em\lower.7ex\hbox{E}\kern-.125emX}}
\begin{document}

\title{Implementation and Analysis of Thermometer Encoding in DWN FPGA Accelerators\\
}

\author{\IEEEauthorblockN{Michael Mecik, Martin Kumm} \\
\IEEEauthorblockA{%
\textit{Fulda University of Applied Sciences, Department of Applied Computer Science},
Fulda, Germany \\
michael.mecik@cs.hs-fulda.de, martin.kumm@cs.hs-fulda.de} \\
}

\maketitle

\begin{abstract}
Fully parallel neural network accelerators on field-programmable gate arrays (FPGAs) offer high throughput for latency-critical applications but face hardware resource constraints. Weightless neural networks (WNNs) efficiently replace arithmetic with logic-based inference. Differential weightless neural networks (DWN) further optimize resource usage by learning connections between encoders and LUT layers via gradient-based training. However, DWNs rely on thermometer encoding, and the associated hardware cost has not been fully evaluated. We present a DWN hardware generator that includes thermometer encoding explicitly. Experiments on the Jet Substructure Classification (JSC) task show that encoding can increase LUT usage by up to 3.20$\times$, dominating costs in small networks and highlighting the need for encoding-aware hardware design in DWN accelerators.
\end{abstract}

\begin{IEEEkeywords}
FPGA, Hardware acceleration, reconfigurable computing, weightless neural networks
\end{IEEEkeywords}

\section{Introduction}

Field-programmable gate arrays (FPGAs) are a compelling choice for accelerating deep neural networks (DNNs) due to their reconfigurable and parallel nature. A wide range of applications, including high-energy physics~\cite{duarte2018fast}, computer vision~\cite{nguyen2019high-throughput, wang2022via, lyu2019chipnet}, speech recognition~\cite{kang2024survey}, natural language processing~\cite{hur2023fast}, network security~\cite{wu2024high-throughput}, and autonomous driving~\cite{li2022fpga}, demand high throughput and low latency. While dataflow architectures enable the acceleration of larger neural networks by processing layers sequentially and mapping them onto dedicated hardware units, fully parallel architectures maximize throughput by executing all operations simultaneously. Practically, they are limited to small or moderately sized neural networks due to the growth in hardware resources required as the model complexity increases, which grows exponentially with input count when neurons are mapped directly to lookup tables (LUTs)~\cite{umuroglu2020logicnets}. Fully parallel architectures can generally be categorized into three classes: (i) architectures that buffer data and share resources, typically for convolutional layers~\cite{tridgell2019unrolling}, to maximize hardware utilization; (ii) LUT-based architectures, where neurons, functions~\cite{andronic2023polylut, andronic2024neuralut, ramirez2025llnn}, or ensembles of simpler neural networks~\cite{weng2025greater, andronic2025neuralut-assemble} are mapped onto the FPGA LUTs, which can be further compressed \cite{cassidy2025reducedlut}; and (iii) weightless neural networks (WNNs), where interconnections are determined during training and no multiplications or accumulations are required~\cite{petersen2022deep, susskind2023uleen, bacellar2024differentiable}.

Recent work has proposed fully parallel WNN architectures known as differential weightless neural networks (DWN), leveraging thermometer encoding and LUT-based logic to avoid arithmetic operations~\cite{bacellar2024differentiable}. 
There are only rare cases where input data is already provided in thermometer-encoded numbers (TEN); usually, analog-to-digital converters or sensors provide positional encoded numbers (PEN), which then require additional conversion to TEN, 
as pointed out in~\cite{andronic2025neuralut-assemble}, this may lead to a ``potentially large overhead in converting into thermometer-encoding''.
To quantify this, we developed a hardware generator for DWN that includes the thermometer encoding stage, with the goal of enabling a complete evaluation of resource usage in fully parallel FPGA deployments.
As the encoding is performed from real-valued input features, quantization becomes necessary, with potential accuracy trade-offs.
Accordingly, we evaluate the best trade-off between accuracy and encoder complexity.

\section{Background}


Some of the most prominent LUT-based architectures include LogicNets~\cite{umuroglu2020logicnets} and its successors, e.g., PolyLUT~\cite{andronic2023polylut}, PolyLUT-Add~\cite{lou2024polylut-add}, and NeuraLUT~\cite{andronic2024neuralut}. Later developments like AmigoLUT~\cite{weng2025greater}, NeuraLUT-Ensemble~\cite{andronic2025neuralut-assemble}, and TreeLUT~\cite{khataei2025treelut} extend this line of work. These architectures heavily rely on pruning and quantization due to scalability challenges, most notably, the exponential hardware cost introduced by increasing LUT fan-in~\cite{umuroglu2020logicnets}.

To increase the functional complexity of individual units mapped to FPGA soft logic, PolyLUT, PolyLUT-Add, and NeuraLUT extend the capabilities of single LUTs. Logic neural networks (LNNs) use discrete logic gates to implement neural-network-like computation. LUT logic-based neural networks (LLNNs) extend this idea by mapping Boolean logic directly onto lookup tables (LUTs). AmigoLUT and NeuraLUT-Ensemble improve model expressiveness by combining ensembles of simpler networks. TreeLUT, on the other hand, introduces a structural innovation by arranging LUTs in a decision tree format.

Unlike LUT-based approaches, WNNs are not sensitive to LUT fan-in and are more efficient. Prior work~\cite{bacellar2024differentiable} demonstrated competitive performance on classification datasets like JSC~\cite{duarte2018fast} and MNIST~\cite{deng2012mnist}, while also showing that mimicking DNN neurons with LUTs may underutilize their capacity. The JSC dataset involves classifying particle collision data into five jet types in high-energy physics, while MNIST focuses on recognizing handwritten digits from 0 to 9. Both serve as benchmarks for fully parallel hardware accelerators.

\begin{figure}[!t]
\centerline{\includegraphics[width=0.43\textwidth]{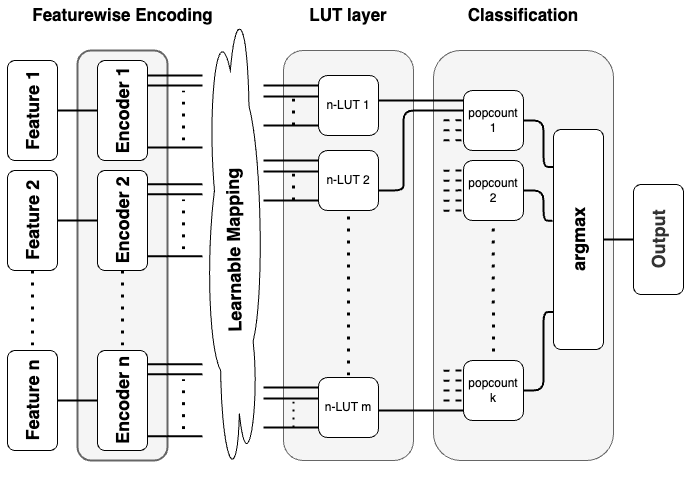}}
\caption{Overview DWN Architecture}
\label{fig:hwgen}
\end{figure}

In~\cite{bacellar2024differentiable}, the model expects thermometer-encoded inputs, LUT layers composed of $m \times n$ input LUTs, and a classification logic based on popcount and argmax operations, as shown in Fig.~\ref{fig:hwgen}. The mapping between the outputs of the thermometer encoders and the LUT layer, and between LUT layers is determined through gradient-based training. For evaluation on the JSC dataset, we use the same model configurations as in~\cite{bacellar2024differentiable}, with a single LUT layer, comprising four variants: \textbf{sm-10}, \textbf{sm-50}, \textbf{md-360}, and \textbf{lg-2400}, where \textbf{sm}, \textbf{md}, and \textbf{lg} denote small, medium, and large models, and the numbers indicate the number of LUTs in the LUT layer. In previous performance evaluations, only the resource usage of the LUT layer and the classification logic was reported, which makes meaningful comparison with LUT-based architectures difficult. We therefore implement the full hardware accelerator, including thermometer encoding, the LUT layer, and the classification logic, to enable a comprehensive evaluation.

\section{Training}

\begin{figure}[!t]
    \scriptsize
    \centering
    \begin{minipage}{0.5\textwidth}
        \centering
        \includegraphics[width=\linewidth]{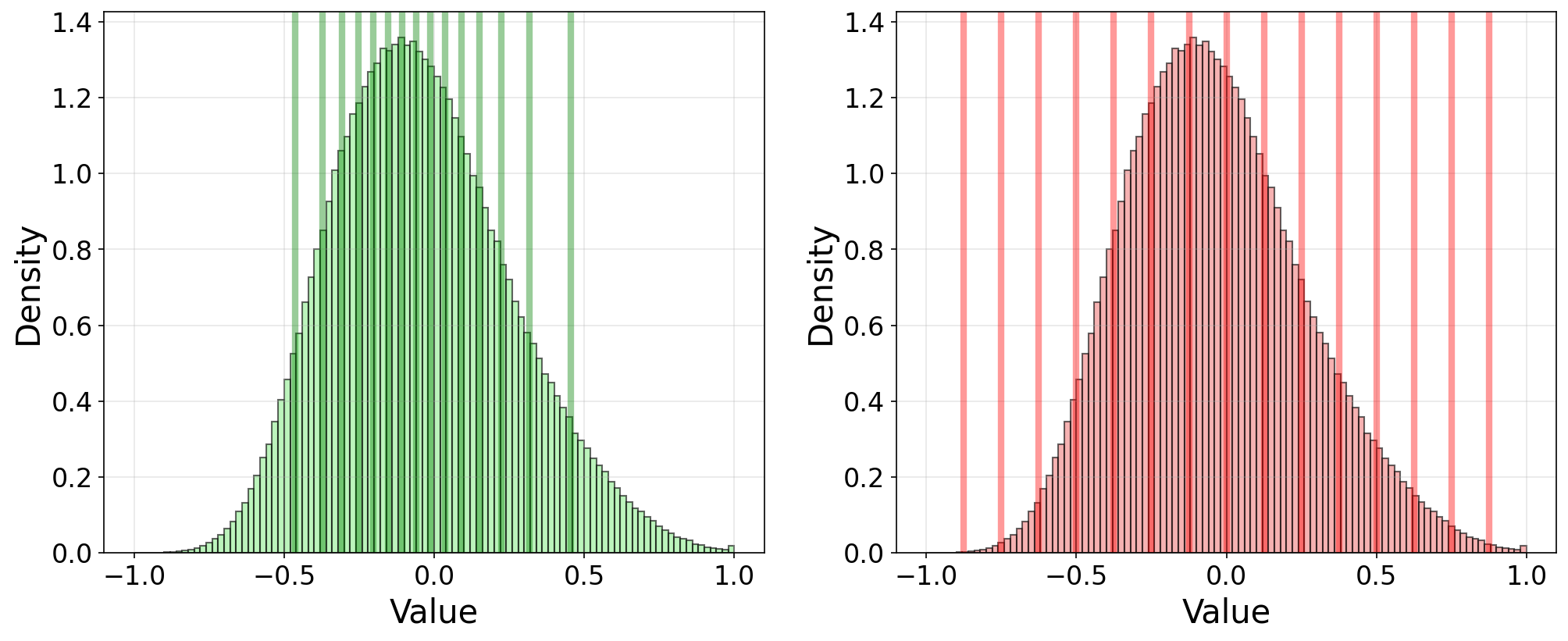}
        \textbf{(a)} JSC Dataset Sample 0 Feature 0
    \end{minipage}
    \hfill
    \begin{minipage}{0.5\textwidth}
        \scriptsize
        \centering
        \includegraphics[width=\linewidth]{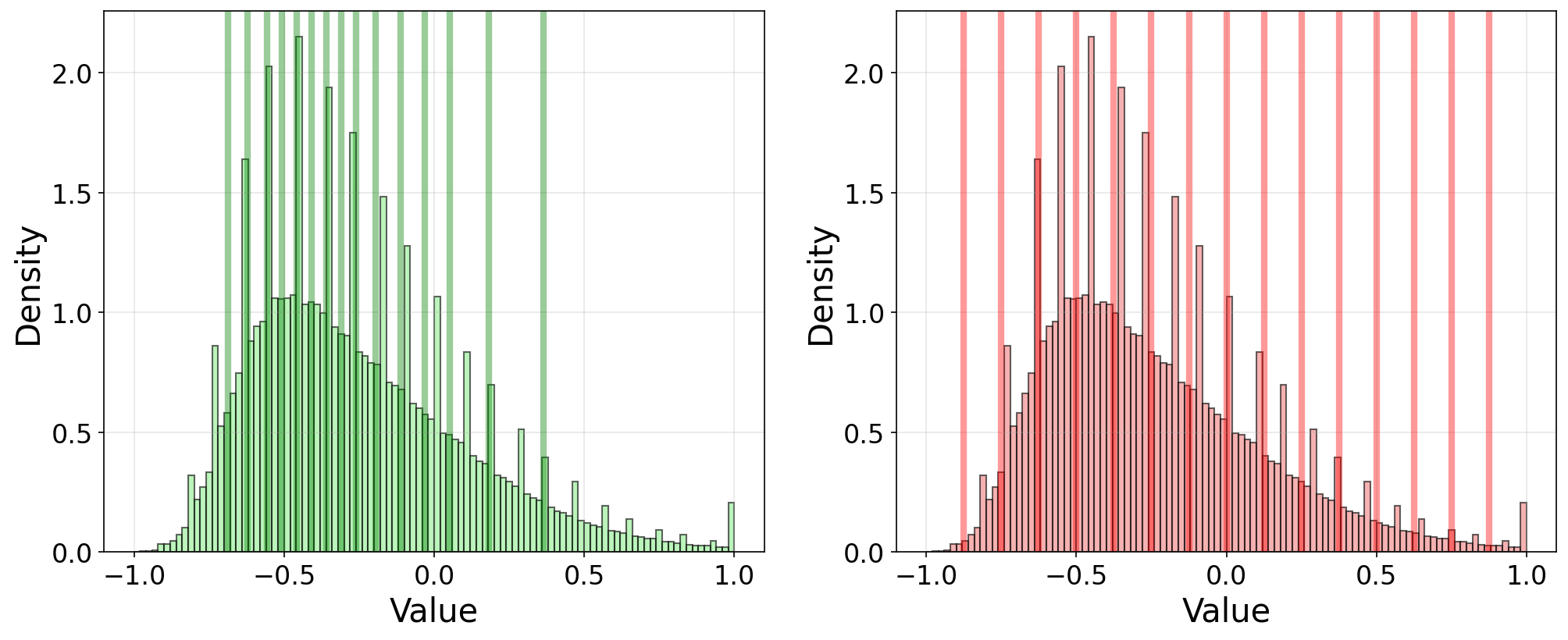}
        \textbf{(b)} JSC Dataset Sample 0 Feature 15
    \end{minipage}
    \caption{Distributive vs. Uniform Encoding of JSC Dataset}
    \label{fig:jsc_thresholds_combined}
\end{figure}

The training procedure for the JSC dataset was adapted from~\cite{bacellar2024differentiable}. 
Prior to training, all input features were normalized to the interval $[-1, 1)$. A non-uniform distributive thermometer encoding was used as introduced in \cite{bacellar2022distributive}, leading to a higher accuracy compared to a uniform decoding. A comparison to uniform encoding is shown in Fig.~\ref{fig:jsc_thresholds_combined} for the example on the first sample of the JSC dataset. To reduce hardware costs associated with the thermometer encoders, the input bit-width was progressively decreased. First, post-training quantization (PTQ) was applied: the thresholds were quantized to a signed fixed-point representation (1,$n$), where the first bit represents the sign and the remaining $n$ bits are fractional. The value of $n$ was determined by progressively reducing the number of fractional bits until the quantized model no longer met its baseline accuracy: 71.1\% for DWN (sm-10), 74.0\% for DWN (sm-50), 75.6\% for DWN (md-360), and 76.3\% for DWN (lg-2400). Models obtained in this manner are denoted as \textbf{DWN-PEN}. To further reduce the input bit-width while maintaining accuracy, fine-tuning was applied to recover performance, resulting in models referred to as \textbf{DWN-PEN+FT}. Each model was fine-tuned using the same procedure. Following post-training quantization (PTQ), the models were trained for 10 epochs using the Adam optimizer with a learning rate of 0.001. A StepLR scheduler was applied with a step size of 30 and a decay factor (gamma) of 0.1 to adjust the learning rate during training.

\section{Implementation}

\begin{figure}[!t]
\centerline{\includegraphics[width=0.20\textwidth]{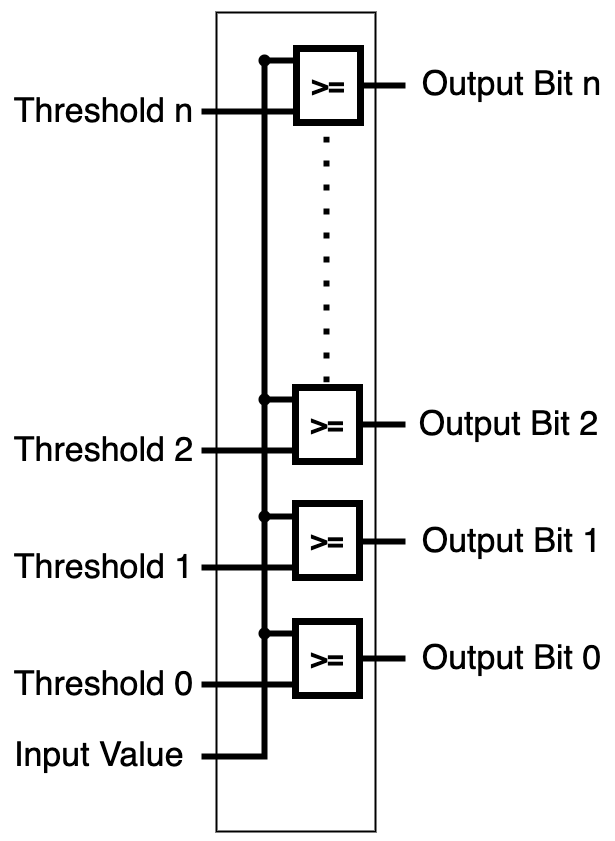}}
\caption{Thermometer Encoder Component}
\label{fig:thresholds}
\end{figure}

\begin{figure}[!t]
\centerline{\includegraphics[width=0.45\textwidth]{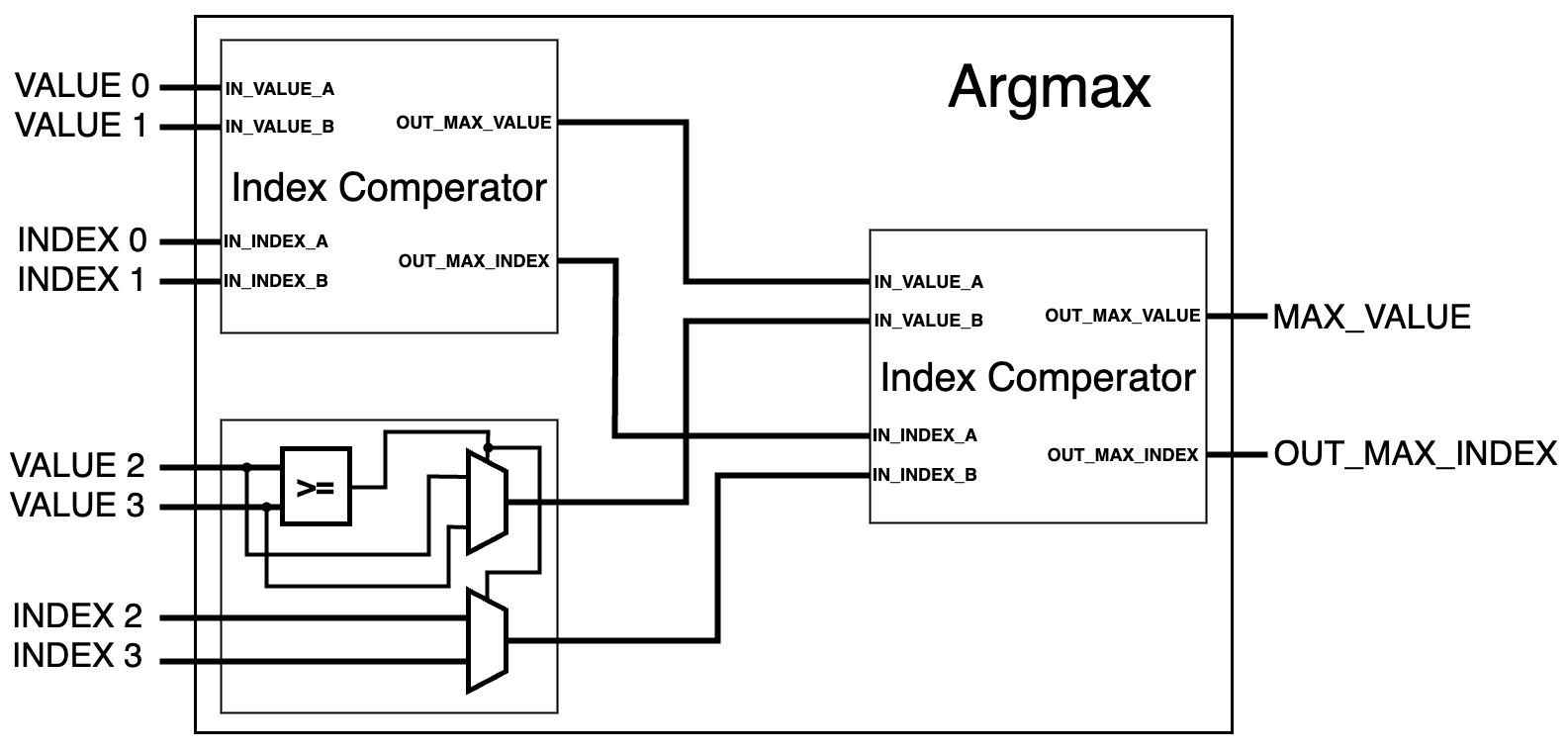}}
\caption{Argmax Component composed of Index Comparators}
\label{fig:argmax}
\end{figure}

\begin{table}[!ht]
\centering
\setlength{\tabcolsep}{2pt} 
\caption{Hardware comparison of DWN-TEN and DWN-PEN+FT for different model sizes.}
\begin{tabular}{lrrrrrr}
\toprule
Model & Acc. & LUT & FF & Fmax & Lat & A$\times$D \\
      & (\%)           &     &    & (MHz) & (ns) & (LUT$\cdot$ns) \\
\midrule
DWN-TEN (lg-2400) & 76.3 & 4,972 & 3,305 & 827 & 7.3 & 36,296 \\
\textbf{DWN-PEN+FT (lg-2400) (9-Bit)} & \textbf{76.3} & \textbf{7,011}& \textbf{961}& \textbf{947}& \textbf{2.1}& \textbf{14,723}\\
\midrule
DWN-TEN (md-360) & 75.6 & 720 & 457 & 827 & 3.6 & 2,592 \\
\textbf{DWN-PEN+FT (md-360) (9-Bit)} & \textbf{75.6} & \textbf{1,697}& \textbf{198}& \textbf{696}& \textbf{2.6}& \textbf{4,412}\\
\midrule
DWN-TEN (sm-50) & 74.0 & 110 & 72 & 1,094 & 1.5 & 165 \\
\textbf{DWN-PEN+FT (sm-50) (8-Bit)} & \textbf{74.0} & \textbf{311}& \textbf{52} & \textbf{1,011}& \textbf{2.0}& \textbf{622}\\
\midrule
DWN-TEN (sm-10) & 71.1 & 20 & 22 & 3,030 & 0.6 & 12 \\
\textbf{DWN-PEN+FT (sm-10) (6-Bit)} & \textbf{71.2} & \textbf{64}& \textbf{18} & \textbf{1,251}& \textbf{1.6}& \textbf{102}\\
\bottomrule
\end{tabular}
\label{tab:dwn-ten-pen}
\end{table}

\begin{table}[!ht]
\centering
\setlength{\tabcolsep}{1.5pt} 
\caption{Comparison of LUT-based Architectures on JSC Dataset}
\begin{tabular}{lrrrrrr}
\toprule
Model & Acc. & LUT & FF & Fmax & Lat & A$\times$D \\
      & (\%) &     &    & (MHz) & (ns) & (LUT$\cdot$ns) \\
\midrule
\textbf{DWN-PEN+FT (lg-2400) (9-Bit)} & \textbf{76.3} & \textbf{7,011}& \textbf{961}& \textbf{947}& \textbf{2.1}& \textbf{14,723}\\
NeuraLUT-Assemble \cite{andronic2025neuralut-assemble}   & 76.0 & 1,780 & 540 & 941 & 2.1 & 3,738 \\
TreeLUT \cite{khataei2025treelut}             & 76.0 & 2,234 & 347 & 735 & 2.7 & 6,032 \\
\midrule
\textbf{DWN-PEN+FT (md-360) (9-Bit)} & \textbf{75.6} & \textbf{1,697}& \textbf{198}& \textbf{696}& \textbf{2.6}& \textbf{4,412}\\
TreeLUT \cite{khataei2025treelut}            & 75.0 & 796 & 74 & 887 & 1.1 & 876 \\
PolyLUT-Add \cite{lou2024polylut-add}        & 75.0 & 36,484 & 1,209 & 315 & 16 & 583,744 \\
NeuraLUT \cite{andronic2024neuralut}           & 75.0 & 92,357 & 4,885 & 368 & 14 & 1,292,998 \\
PolyLUT \cite{andronic2023polylut}            & 75.0 & 236,541 & 2,775 & 235 & 21 & 4,967,361 \\
LLNN \cite{ramirez2025llnn}               & 75.0 & 13,926 & 0 & 153 & 6.5 & 90,519 \\
\midrule
ReducedLUT \cite{cassidy2025reducedlut}             & 74.9 & 58,409 & 0 & 303 & 17  & 992,963 \\
AmigoLUT-NeuraLUT-S \cite{weng2025greater}    & 74.4 & 42,742 & 4,717 & 520 & 9.6 & 410,323 \\
\textbf{DWN-PEN+FT (sm-50) (8-Bit)} & \textbf{74.0} & \textbf{311}& \textbf{52} & \textbf{1,011}& \textbf{2.0}& \textbf{622}\\
LogicNets* \cite{umuroglu2020logicnets}             & 73.1 & 36,415 & 2,790 & 390 & 6   & 218,490 \\
\midrule
AmigoLUT-NeuraLUT-XS \cite{weng2025greater} & 72.9 & 1,243 & 1,240 & 1,008 & 5.0 & 6,215 \\
ReducedLUT \cite{cassidy2025reducedlut}          & 72.5 & 2,786 & 0    & 409  & 4.9 & 13,651 \\
LogicNets* \cite{umuroglu2020logicnets}          & 72.1 & 15,526 & 881  & 577 & 5   & 77,630 \\
PolyLUT \cite{andronic2023polylut}             & 72.0 & 12,436 & 773  & 646 & 5   & 62,180 \\
NeuraLUT \cite{andronic2024neuralut}            & 72.0 & 4,684  & 341  & 727 & 3   & 14,148 \\
PolyLUT-Add \cite{lou2024polylut-add}         & 72.0 & 895   & 189  & 750 & 4   & 3,580 \\
LLNN \cite{ramirez2025llnn}                & 72.0 & 6,431  & 0    & 449 & 2.2 & 14,148 \\
\midrule
\textbf{DWN-PEN+FT (sm-10) (6-Bit)} & \textbf{71.2} & \textbf{64}& \textbf{18} & \textbf{1,307}& \textbf{1,6}& \textbf{102}\\
AmigoLUT-NeuraLUT-XS \cite{weng2025greater} & 71.1 & 320 & 482 & 1,445 & 3.5 & 1,120 \\
\bottomrule
\end{tabular}
\begin{tablenotes}[flushleft]
\vspace{2pt}   
\footnotesize
\item * Updated LogicNets results from https://github.com/Xilinx/logicnets
\end{tablenotes}
\label{tab:lut-based-implementations}
\end{table}

\begin{table}[!ht]
\centering
\setlength{\tabcolsep}{2pt} 
\caption{Comparison of DWN variants (TEN, PEN, PEN+FT) on JSC, showing accuracy, LUTs, bit-width (BW).}
\begin{tabular}{lcccccccccc}
\toprule
\multirow{2}{*}{Model} &
\multicolumn{3}{c}{PEN+FT} &
\multicolumn{3}{c}{PEN} &
\multicolumn{2}{c}{TEN} \\
\cmidrule(lr){2-4}
\cmidrule(lr){5-7}
\cmidrule(lr){8-9}
& Acc. & LUTs & BW & Acc. & LUTs & BW & Acc. & LUTs \\
\midrule
DWN (sm-10)   & 71.2 & 64 (+320\%) & 6  & 71.3 & 106 (+530\%) & 9 & 71.1 & 20   \\
DWN (sm-50)   & 74.0 & 311 (+283\%)& 8 & 74.0 & 345 (+313\%) & 9  & 74.0 & 110  \\
DWN (md-360)  & 75.6 & 1,697 (+236\%)& 9  & 75.6 & 1,994 (+277\%) & 11 & 75.6 & 720  \\
DWN (lg-2400) & 76.3 & 7,011 (+141\%) & 9  & 76.3 & 18,330 (+368\%) & 12 & 76.3 & 4,972 \\
\bottomrule
\end{tabular}
\label{tab:ten-pen-pen-ft}
\end{table}

\begin{figure}[!t]
    \centerline{\includegraphics[width=0.4\textwidth]{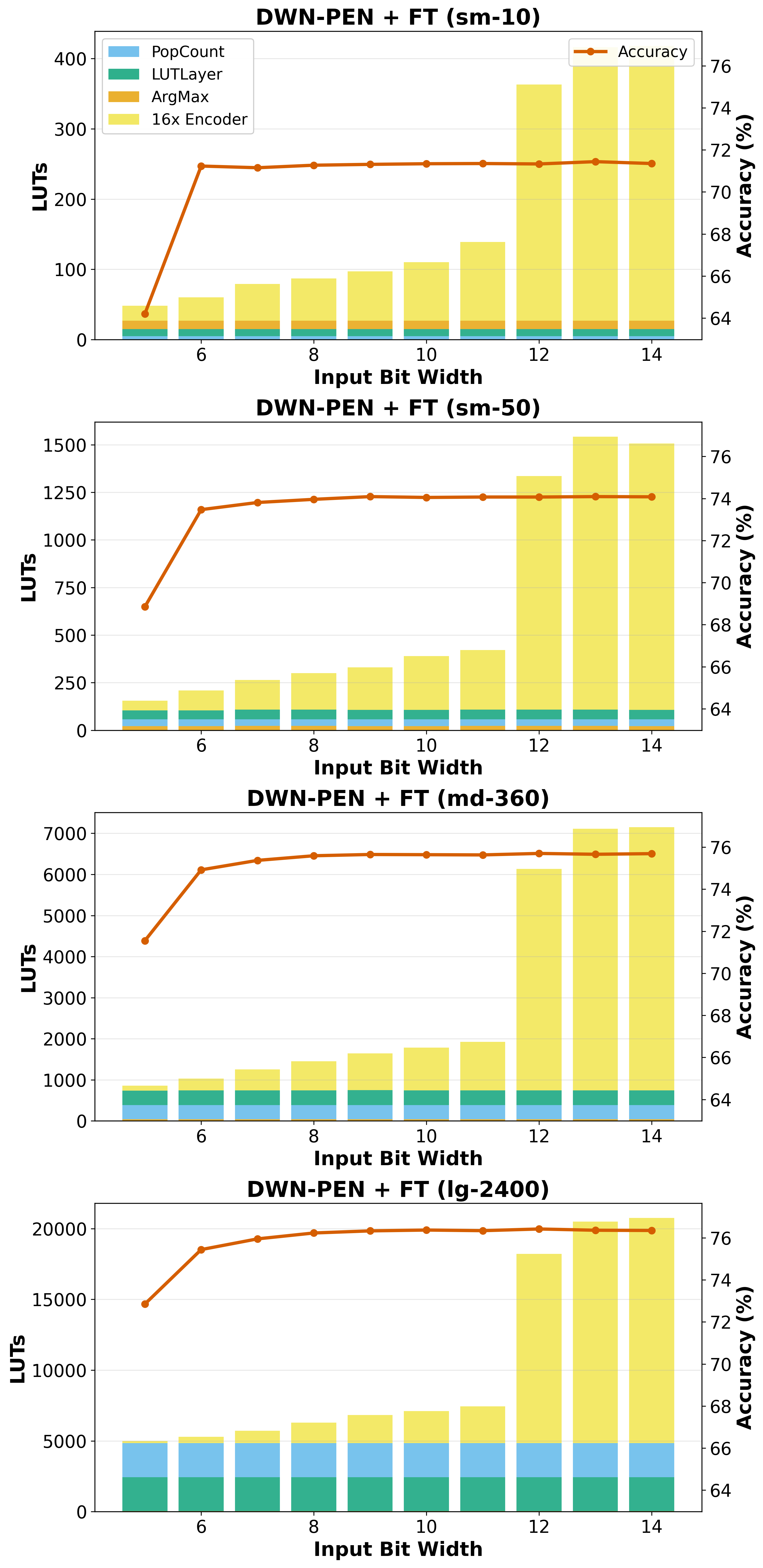}}
    \caption{Component Breakdown for DWN-PEN + FT}
    \label{fig:dwn-pen-ft}
\end{figure}

\begin{figure*}[!t]
\centerline{\includegraphics[width=0.63\textwidth]{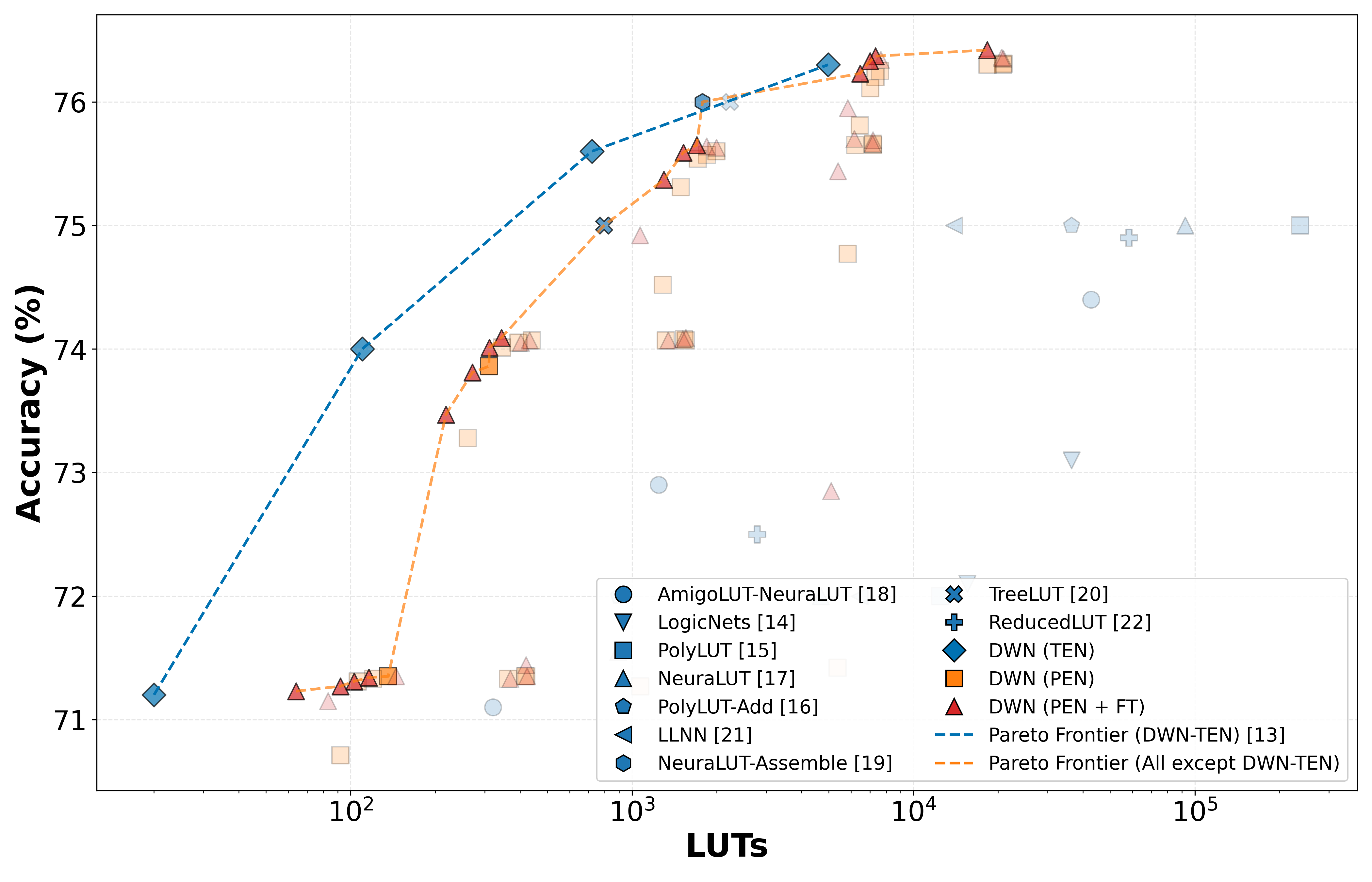}}
\caption{Pareto-Frontier LUT-based Architectures for JSC} 
\label{fig:pareto}
\end{figure*}

To evaluate the impact of thermometer encoding, a hardware generator was developed to target the DNN architecture from~\cite{bacellar2024differentiable}. Core components of the DWN architecture, the thermometer encoder, LUT layer, and classification logic, were implemented using the FloPoCo framework~\cite{de_dinechin2024application}. The thermometer encoder and LUT layer were developed from scratch, while the popcount operation (based on compressor trees~\cite[p.~153--156]{de_dinechin2024application}) was reused from FloPoCo.. In distributive encoding (percentile-based thresholding), the non-uniform threshold values require each threshold to be evaluated independently, necessitating a separate comparator for every threshold level as shown in Fig. \ref{fig:thresholds}. This contrasts with uniform encoding, where evenly spaced thresholds allow for a simpler hardware structure. In our design, we implement distributive encoding to benefit from its improved accuracy despite the additional hardware complexity. The mapping between thermometer encoder outputs and the LUT layer was derived directly from the trained software model. Both the LUT layer and the classification logic were fully implemented, and the LUT utilization for JSC-based models reported in \cite{bacellar2024differentiable} was successfully reproduced. The argmax module compares the popcount values of all classes and outputs the maximum value and its index. 
For example, in the JSC dataset, which has 5 output classes, the DWN (sm-50) configuration uses a LUT layer with 50 LUTs. Since each class is represented by 10 LUTs, each group of 10 LUTs forms one popcount value, which serves as the input to the argmax stage. In this stage, the comparators evaluate the inputs pairwise, propagating the larger popcount value at each step. If two inputs have the same popcount value, the class with the lower index is selected. This reduction continues until all classes have been processed and only the final maximum value and its index remain. Figure~\ref{fig:argmax} shows how the argmax operation processes four inputs to determine the maximum value and its corresponding index.

\section{Results}

All results were produced on the same device as prior work (AMD \texttt{xcvu9p-flgb2104-2-i}) for fair comparison. We followed the same methodology, performing synthesis in \texttt{out-of-context mode (OOC)} using Vivado’s \texttt{Flow\_PerfOptimized\_high} strategy and operating at a clock frequency of 700 MHz. In Table \ref{tab:dwn-ten-pen}, the TEN and PEN+FT implementations of DWN for the JSC datasets are compared. As expected, models including thermometer encoding are significantly larger than models that use ternary encoded values. The accuracy of the individual models with and without PEN is nearly identical, but frequency and latency vary significantly. 

Compared to other state-of-the-art LUT-based architectures, DWN models after PTQ and fine-tuning demonstrate competitive hardware utilization, as shown in Table~\ref{tab:lut-based-implementations}. Small models, such as DWN (sm-10) and DWN (sm-50), achieve the highest efficiency in the accuracy range of 71.1\%–74.0\%. In the range $75.0\% \leq \text{accuracy} < 76.0\%$, TreeLUT is more resource-efficient, although DWN still attains higher accuracy. For accuracies above 76.0\%, NeuraLUT-Assemble outperforms both TreeLUT and DWN resource-wise. Overall, DWN achieves the highest accuracy among the compared architectures but requires more than three times the hardware resources compared to NeuraLUT-Assemble.

As previously mentioned, quantization and fine-tuning were applied to further reduce hardware costs. For the smallest model, DWN (sm-10), the gap between PEN and TEN decreased from 5.30× to 3.20×. For the largest model, DWN (lg-2400), the hardware costs were reduced from 3.68× down to 1.41×, narrowing even the difference between DWN-PEN and DWN-TEN as shown in Table \ref{tab:ten-pen-pen-ft}.

A component breakdown for the DWN-PEN+FT models is shown in Fig. \ref{fig:dwn-pen-ft}, along with the corresponding accuracy for each input bit-width. For smaller models such as DWN (sm-10), DWN (sm-50), and DWN (md-360), the thermometer encoders dominate the overall hardware costs, even at low input bit-widths. For larger models such as DWN (lg-2400), the encoder cost becomes less dominant, particularly for input bit-widths below 10 bits, at which point the LUT layer and popcount components contribute the most.

Fig. \ref{fig:pareto} shows a Pareto plot of LUT-based architectures. DWN-TEN forms its own front, highlighting the original results \cite{bacellar2024differentiable}, while DWN-PEN and DWN-PEN+FT perform very well even with integrated thermometer encoding. NeuraLUT-Assemble and TreeLUT lie above the DWN-TEN front and scale better in the high-accuracy region.

\section{Conclusion}

In this work, we analyzed the hardware cost of integrating thermometer encoding into DWNs using the hardware generator described earlier. This setup allowed us to isolate and quantify the hardware contributions of each component across different input bit-widths and network scales.

Our results show that thermometer encoding increases LUT usage by up to 5.30× at 9-bit inputs without fine-tuning for DWN (sm-10). With fine-tuning, the input bit-width can be reduced to 6 bits while maintaining nearly the same accuracy, lowering the overhead to 3.20×. For large models such as DWN (lg-2400), fine-tuning reduces the input bit width from 11 to 9 bits, making the PEN variant only 1.41× larger than the TEN variant. At reduced input bit-widths, hardware utilization is largely dominated by the popcount and LUT layers.

Overall, thermometer encoding introduces overhead, but this can be substantially reduced through input quantization and fine-tuning, allowing DWNs to maintain competitive accuracy and hardware utilization. While this work focused primarily on optimization via input quantization, the size of the thermometer outputs might also significantly affect hardware cost. In the proposed model setup, each thermometer encoder produces 200 output bits per feature; for the JSC dataset with 16 features, this results in 3,200 bits, which can become costly especially for more complex datasets.

Future work could focus on hardware-software co-design to further reduce hardware costs and improve efficiency: (i) evaluating and reducing thermometer encoder outputs by decreasing the number of bits per feature, for example, a LUT layer with 50 LUTs may require 300 or fewer input connections (50 × 6-LUT), whereas for the JSC dataset with 16 thermometer encoders, 3,200 outputs are currently provided, (ii) reducing the number of LUTs in the LUT layer through pruning, which also reduces popcount complexity, (iii) exploring mixed-precision quantization for inputs, which would involve evaluating which features contribute more or less to model performance and using simpler threshold schemes, such as uniform thermometer encoding, and (iv) optimizing the classification logic, since for large models such as DWN (lg-2400), the popcount and LUT layers dominate hardware utilization at smaller input bit-widths. Addressing these points together would benefit from a co-design approach that jointly considers network architecture, encoding schemes, and hardware implementation.
\FloatBarrier

\vspace{12pt}

\begin{thebibliography}{00}
\bibitem{duarte2018fast}Duarte, J., Han, S., Harris, P., Jindariani, S., Kreinar, E., Kreis, B., Ngadiuba, J., Pierini, M., Rivera, R., Tran, N. \& Others Fast inference of deep neural networks in FPGAs for particle physics. {\em Journal Of Instrumentation}. \textbf{13}, P07027 (2018)


\bibitem{nguyen2019high-throughput}Nguyen, D., Nguyen, T., Kim, H. \& Lee, H. A high-throughput and power-efficient FPGA implementation of YOLO CNN for object detection. {\em IEEE Transactions On Very Large Scale Integration (VLSI) Systems}. \textbf{27}, 1861-1873 (2019)
\bibitem{wang2022via}Wang, T., Gong, L., Wang, C., Yang, Y., Gao, Y., Zhou, X. \& Chen, H. ViA: A Novel Vision-Transformer Accelerator Based on FPGA. {\em IEEE Transactions On Computer-Aided Design Of Integrated Circuits And Systems}. \textbf{41}, 4088-4099 (2022)
\bibitem{lyu2019chipnet}Lyu, Y., Bai, L. \& Huang, X. ChipNet: Real-Time LiDAR Processing for Drivable Region Segmentation on an FPGA. {\em IEEE Transactions On Circuits And Systems I: Regular Papers}. \textbf{66}, 1769-1779 (2019)


\bibitem{kang2024survey}Kang, B., Lee, H., Yoon, S., Kim, Y., Jeong, S., Kim, H. \& Others A survey of FPGA and ASIC designs for transformer inference acceleration and optimization. {\em Journal Of Systems Architecture}. pp. 103247 (2024)


\bibitem{hur2023fast}Hur, S., Na, S., Kwon, D., Kim, J., Boutros, A., Nurvitadhi, E. \& Kim, J. A fast and flexible fpga-based accelerator for natural language processing neural networks. {\em ACM Transactions On Architecture And Code Optimization}. \textbf{20}, 1-24 (2023)


\bibitem{wu2024high-throughput}Wu, M. \& Kondo, M. A High-Throughput Network Intrusion Detection System Using On-Device Learning on FPGA. {\em 2024 IEEE 17th International Symposium On Embedded Multicore/Many-core Systems-on-Chip (MCSoC)}. pp. 426-433 (2024)
\bibitem{deng2012mnist}Deng, L. The mnist database of handwritten digit images for machine learning research. {\em IEEE Signal Processing Magazine}. \textbf{29}, 141-142 (2012)



\bibitem{li2022fpga}Li, Y., Li, S., Jia, X., Zeng, S. \& Wang, Y. FPGA accelerated model predictive control for autonomous driving. {\em Journal Of Intelligent And Connected Vehicles}. \textbf{5}, 63-71 (2022)

\bibitem{tridgell2019unrolling}Tridgell, S., Kumm, M., Hardieck, M., Boland, D., Moss, D., Zipf, P. \& Leong, P. Unrolling Ternary Neural Networks.  (2019,12)


\bibitem{petersen2022deep}Petersen, F., Borgelt, C., Kuehne, H. \& Deussen, O. Deep differentiable logic gate networks. {\em Advances In Neural Information Processing Systems}. \textbf{35} pp. 2006-2018 (2022)

\bibitem{susskind2023uleen}Susskind, Z., Arora, A., Miranda, I., Bacellar, A., Villon, L., Katopodis, R., Araújo, L., Dutra, D., Lima, P., França, F., Breternitz Jr., M. \& John, L. ULEEN: A Novel Architecture for Ultra-low-energy Edge Neural Networks. {\em ACM Trans. Archit. Code Optim.}. \textbf{20} (2023,12), https://doi.org/10.1145/3629522

\bibitem{bacellar2024differentiable}Bacellar, A., Susskind, Z., Breternitz Jr, M., John, E., John, L., Lima, P. \& França, F. Differentiable weightless neural networks. {\em ArXiv Preprint ArXiv:2410.11112}. (2024)

\bibitem{umuroglu2020logicnets}Umuroglu, Y., Akhauri, Y., Fraser, N. \& Blott, M. LogicNets: Co-Designed Neural Networks and Circuits for Extreme-Throughput Applications. {\em 2020 30th International Conference On Field-Programmable Logic And Applications (FPL)}. pp. 291-297 (2020)

\bibitem{andronic2023polylut}Andronic, M. \& Constantinides, G. PolyLUT: learning piecewise polynomials for ultra-low latency FPGA LUT-based inference. {\em 2023 International Conference On Field Programmable Technology (ICFPT)}. pp. 60-68 (2023)

\bibitem{lou2024polylut-add}Lou, B., Rademacher, R., Boland, D. \& Leong, P. PolyLUT-Add: FPGA-based LUT Inference with Wide Inputs. {\em 2024 34th International Conference On Field-Programmable Logic And Applications (FPL)}. pp. 149-155 (2024)

\bibitem{andronic2024neuralut}Andronic, M. \& Constantinides, G. NeuraLUT: Hiding neural network density in boolean synthesizable functions. {\em 2024 34th International Conference On Field-Programmable Logic And Applications (FPL)}. pp. 140-148 (2024)

\bibitem{weng2025greater}Weng, O., Andronic, M., Zuberi, D., Chen, J., Geniesse, C., Constantinides, G., Tran, N., Fraser, N., Duarte, J. \& Kastner, R. Greater than the Sum of its LUTs: Scaling Up LUT-based Neural Networks with AmigoLUT. {\em Proceedings Of The 2025 ACM/SIGDA International Symposium On Field Programmable Gate Arrays}. pp. 25-35 (2025)

\bibitem{andronic2025neuralut-assemble}Andronic, M. \& Constantinides, G. NeuraLUT-Assemble: Hardware-aware Assembling of Sub-Neural Networks for Efficient LUT Inference. {\em ArXiv Preprint ArXiv:2504.00592}. (2025)

\bibitem{khataei2025treelut}Khataei, A. \& Bazargan, K. TreeLUT: An Efficient Alternative to Deep Neural Networks for Inference Acceleration Using Gradient Boosted Decision Trees. {\em Proceedings Of The 2025 ACM/SIGDA International Symposium On Field Programmable Gate Arrays}. pp. 14-24 (2025)


\bibitem{ramirez2025llnn}Ramirez, I., Garcia-Espinosa, F., Concha, D., Aranda, L. \& Schiavi, E. LLNN: A Scalable LUT-Based Logic Neural Network Architecture for FPGAs. {\em IEEE Transactions On Circuits And Systems I: Regular Papers}. (2025)

\bibitem{cassidy2025reducedlut}Cassidy, O., Andronic, M., Coward, S. \& Constantinides, G. ReducedLUT: Table Decomposition with "Don't Care" Conditions. {\em Proceedings Of The 2025 ACM/SIGDA International Symposium On Field Programmable Gate Arrays}. pp. 36-42 (2025), https://doi.org/10.1145/3706628.3708823

\bibitem{bacellar2022distributive}Bacellar, A., Susskind, Z., Villon, L., Miranda, I., Araújo, L., Dutra, D., Breternitz Jr, M., John, L., Lima, P. \& França, F. Distributive thermometer: A new unary encoding for weightless neural networks. {\em ESANN 2022 Proceedings}. (2022)

\bibitem{de_dinechin2024application}Dinechin, F. \& Kumm, M. Application-Specific Arithmetic. (Springer,2024)



\end{thebibliography}
\end{document}